\documentclass [12pt]{article}

\usepackage{graphicx}   % allow inclusion of postscript figures

\usepackage{amsmath}     % AMS math packages

\usepackage{amssymb}     % AMS symbols

\usepackage{amsfonts}    % AMS fonts

\begin{document}

%\begin{singlespace}    

\begin{center}

{\bf {\LARGE Quiescent Cosmology and the Final State}}

\vspace {1 mm}

{\bf {\LARGE of the Universe}}

\vspace {3 cm}

{\bf Philipp A.\ H\"ohn}\footnote {Email address: p.a.hohn@uu.nl} and {\bf Susan M.\ Scott}\footnote {Email address: susan.scott@anu.edu.au}

\vspace {0.5 cm}

\emph{Centre for Gravitational Physics,}

\emph{Department of Physics,}

\emph{College of Science,}

\emph{The Australian National University}

\emph{Canberra ACT 0200}

\emph{AUSTRALIA}

\vspace {2 cm}

{\bf Abstract}\footnote {This essay was written expressly for the Gravity Research Foundation 2007 Awards for Essays on Gravitation.}

\end{center}

{\small It has long been a primary objective of cosmology to understand the apparent isotropy in our universe and to provide a mathematical formulation for its evolution. A school of thought for its explanation is \emph{quiescent cosmology}, which already possesses a mathematical framework, namely the definition of an \emph{isotropic singularity}, but only for the initial state of the universe. A complementary framework is necessary in order to also describe possible final states of the universe. Our new definitions of an \emph{anisotropic future endless universe} and an \emph{anisotropic future singularity}, whose structure and properties differ significantly from those of the \emph{isotropic singularity}, offer a promising realisation for this framework. The combination of the three definitions together may then provide the first complete formalisation of the \emph{quiescent cosmology} concept.}

%\end{singlespace}    

%\vspace{2cm}
\newpage

Dating from the early examinations of the cosmic microwave background in the 1960s \cite{Penzias}, it has been well known that the energy distribution in the observable vicinity of our universe is highly isotropic around us. The explanation and mathematical formalisation of the evolution of this isotropy has long been a primary objective of cosmology, stimulating the development of competing schools of thought.

The first prominent attempt to explain the isotropy was Misner's \emph{chaotic cosmology} \cite{Misner} which is based on the idea that the universe originated in a highly irregular and chaotic beginning. The isotropy was conjectured to have been produced through dissipative effects, such as  particle creation or collisions, and is seen today because we simply happen to live at a somewhat \emph{late} stage of the evolution of our universe. This idea was attractive since a detailed knowledge about the exact initial state of the universe was not necessary in order to understand the current one, as any microstate of a maximally chaotic beginning would lead to the universe as we now observe it. \emph{Chaotic cosmology}, however, seems untenable in its full generality, as detailed analyses have indicated \cite{Collins, Hawking, Barrow1977, Penrose1979}.

Instead, the concept of \emph{quiescent cosmology} was established by Barrow as an alternative explanation \cite{Barrow1978}. In direct contrast to the previous scenario, it was proposed that the universe originated in a very smooth and regular beginning, but evolved away from regularity because of the attractive nature of gravity which becomes dominant on large scales. A classical cosmological model describing our own universe must thereby possess an initial singularity similar to the one in the isotropic FRW models. The apparent isotropy observed today would thus be due to the fact that we must live at a somewhat \emph{early} stage of the evolution of our universe.

Penrose provided a possible justification for the initial isotropy of the universe by introducing the notion of \emph{gravitational entropy} \cite{Penrose1979}, which measures the contribution of the gravitational field to the total entropy and behaves somewhat anomalously in that it increases with the degree of clumping of matter and becomes maximal in a black hole - which might seem counter-intuitive at first sight when one considers the usual behaviour of entropy with regard to the matter distribution. This behaviour ows to the fact that the gravitational force is (on the classical level) universally attractive. He reasoned that, due to the arrow of time, the initial state of the universe must, in some way, have been a low-entropy one, and associated this low-entropy with a low degree of clumping,  corresponding to isotropy. By relating a low degree of clumping with a bounded Weyl curvature, Penrose drew a connection between the Weyl tensor and gravitational entropy, and arrived at what is now known as the \emph{Weyl Curvature Hypothesis} (WCH): the expectation that the Weyl curvature would initially be matter (i.e.\ Ricci) dominated and that the converse would be true at a cosmological future. The simplest and most frequently used method \cite{GW1985, GCW1992} of probing the WCH is to examine the scalar
\begin{eqnarray}
K\equiv\frac{C_{abcd}C^{abcd}}{R_{ef}R^{ef}}.
\end{eqnarray}

In order to fully investigate the implications of these ideas, it is necessary to devise a suitable framework for \emph{quiescent cosmology} to encode these ideas mathematically. The geometric definition of an \emph{Isotropic Singularity} by Goode and Wainwright \cite{GW1985}---which we will henceforth call an \emph{Isotropic Past Singularity} (IPS)---beautifully encapsulates and generalises a large amount of previous work on initial cosmological singularities and provides an appropriate framework, at least for the initial state.

The definition of an IPS employs a conformal structure and relates the geometry of the physical universe $(\mathcal{M},\mathbf{g})$ to that of an unphysical space-time $(\tilde{\mathcal{M}},\tilde{\mathbf{g}})$ via a conformal factor $\Omega(T)$, which merely depends on a cosmic time function $T$ defined on $\tilde{\mathcal{M}}$ (where $\mathcal{M}\subset\tilde{\mathcal{M}}$), i.e.
\begin{eqnarray}
\mathbf{g}=\Omega^{2}\left(T\right)\tilde{\mathbf{g}},
\end{eqnarray}
where entities equipped with a $\tilde{} \ $ relate to the unphysical space-time. The elegance of the definition of an IPS, which is depicted in Fig.\ 1, lies in the fact that the entire singular behaviour of the physical universe is encoded in the vanishing of $\Omega$ at the cosmic time $T_i$ of the big bang, while the unphysical space-time remains completely regular on an open neighbourhood of the slice $T_i$, thereby greatly simplifying calculations and investigations.

With the cosmological fluid congruence $\mathbf{u}$ in $\mathcal{M}$ can be associated a timelike congruence $\tilde{\mathbf{u}}$ in $\tilde{\mathcal{M}}$ satisfying
\begin{eqnarray}
\tilde{{\mathbf{u}}}=\Omega\mathbf{u} \ \ \text{in} \ \ \mathcal{M}.
\end{eqnarray}
Using this framework, one can then prove that a cosmology with an IPS does indeed behave very isotropically in the beginning, with an initially vanishing $K$ and expansion dominated kinematics, in complete accordance with \emph{quiescent cosmology} \cite{GW1985}. 

\begin{figure}

\centering

\includegraphics[width=8.5cm, height=7cm]{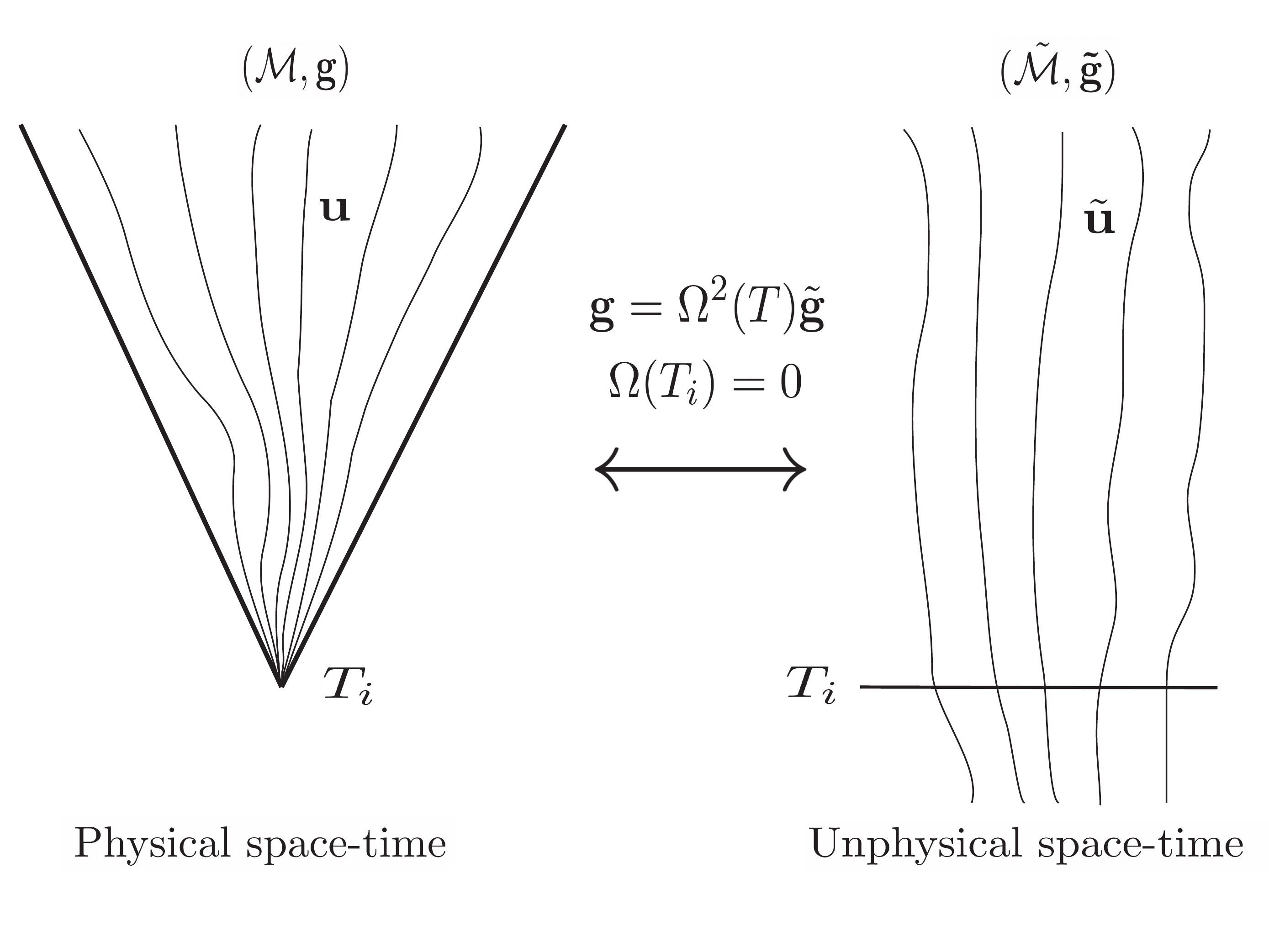}

%\end{figure}

%  	\inclimg{coolingfct}{}

%\begin{center}

%\vspace*{5cm}

\caption{{\footnotesize The schematic interpretation of the definition of an IPS }}

%\end{center}

\end{figure}

 There are many cosmological models which are known to admit an IPS \cite{ScottEricksson1998} and much progress has been made in the quest to find the precise and necessary conditions, in general, for an IPS to occur \cite{Erickssonthesis}. One can, moreover, argue that \emph{quiescent cosmology}, once formulated in the framework of an IPS, is a viable alternative to the idea of cosmic inflation \cite{GCW1992}. It is not necessarily the case, however, that an appropriate future behaviour of a cosmological model is guaranteed by the existence of an IPS.

We need to complete the formalisation of \emph{quiescent cosmology} to include possible final states of our universe. Projecting \emph{quiescent cosmology} and the WCH forward in time, it is clear that the local degree of clumping, and therefore the local anisotropy, in the universe must increase if the cosmological future is to be a high-entropy state. Since FRW universes are isotropic throughout and a subclass of them admit an IPS, it becomes clear that the definition of Goode and Wainwright is not sufficient, in itself, to guarantee a future evolution of the respective model which is consistent e.g.\ with the WCH. A further framework is necessary in order to complete the picture of the IPS in an appropriate way. Since neither a re-collapsing nor an ever-expanding universe can be ruled out by our current knowledge of the universe, we would like to find new definitions covering both scenarios.

The recently proposed geometric definitions of an \emph{Anisotropic Future Endless Universe} (AFEU) and an \emph{Anisotropic Future Singularity} (AFS) \cite{ScottHoehn1} are the promising results of an analysis guided by example cosmologies and general considerations. Similarly to the case of the IPS, both employ conformal structures, such as 
\begin{eqnarray}
\mathbf{g}=\bar{\Omega}^{2}\left(\bar{T}\right)\bar{\mathbf{g}}, \;\; \bar{\mathbf{u}}=\bar{\Omega}\mathbf{u},
\end{eqnarray}
where entities in the unphysical space-time $(\mathcal{M},\bar{\mathbf{g}})$ as well as the conformal factor $\bar{\Omega}$ and the cosmic time function $\bar{T}$ itself are now equipped with a $\bar{} \ $ to avoid confusion with the respective entities at the IPS.

Our work proves that, unlike in the case of the IPS, a regular unphysical space-time at the cosmic time function value $\bar{T}_f$ of the future infinity or singularity is no longer compatible with the WCH; one can show that it necessarily leads to a future asymptotic isotropy, namely a vanishing $K$ and expansion dominated kinematics, independently of whether the conformal factor vanishes or diverges at $\bar{T}_f$  \cite{ScottHoehn1}. Instead, it is therefore necessary for the unphysical space-time to possess irregularities, such as a metric degeneracy at $\bar{T}_f$, if an asymptotic future anisotropy\footnote{For example, Weyl-dominated Ricci curvature and kinematics which are not dominated by the (isotropic) expansion $\theta$.} is required to occur. The important conclusion of this state of affairs is that \emph{if we are to follow the ideas of \emph{quiescent cosmology} and the WCH employing  conformal structures involving a conformal factor as a function of cosmic time, then a regular conformal metric for the classical initial state of the universe necessarily produces the desired asymptotic initial isotropy, while we should require a conformal structure with irregular conformal metric for the cosmological future}.

The definitions of an AFS and an AFEU, consequently, possess an unphysical metric which remains finite, but becomes degenerate as $\bar{T}\rightarrow\bar{T}_f$. Moreover, motivated by example cosmologies and in contrast to the definition of an IPS, the new conformal factor $\bar{\Omega}(\bar{T})$ now diverges at the AFS and in the AFEU, i.e.\ the divergent behaviour of the physical metric is entirely encoded in $\bar{\Omega}(\bar{T})$. The known example cosmologies for the AFEU are the Carneiro-Marugan models, a subclass of the Szekeres models and the Kantowski models; the only known models which admit an AFS are the Kantowski-Sachs models \cite{ScottHoehn1,ScottHoehn2}. The two new definitions only differ in the behaviour of the determinant of the physical metric which must vanish at the AFS and diverge in the AFEU. Since the new definitions do not require any stringent assumptions about the initial state of the universe, they admit the great advantage of also being compatible with inflationary cosmologies leading to future anisotropies. Their pictorial interpretations are given in Fig.\ 2 and Fig.\ 3.

\begin{figure}

\centering

\includegraphics[width=8.5cm, height=7cm]{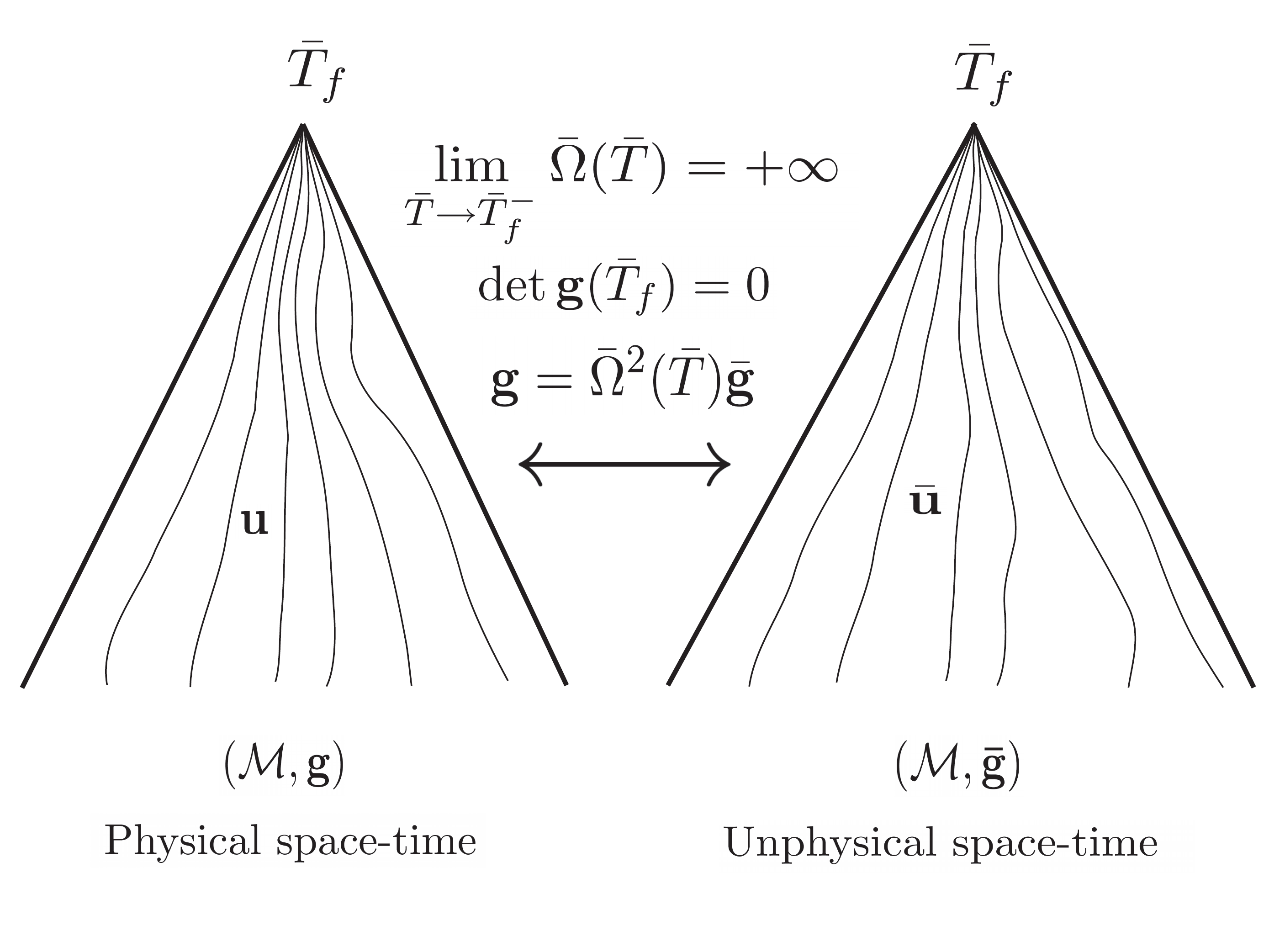}

%\end{figure}

%  	\inclimg{coolingfct}{}

%\begin{center}

%\vspace*{5cm}

\caption{{\footnotesize The schematic interpretation of the definition of an AFS }}

%\end{center}

\end{figure}

\begin{figure}

\centering

\includegraphics[width=8.5cm, height=7cm]{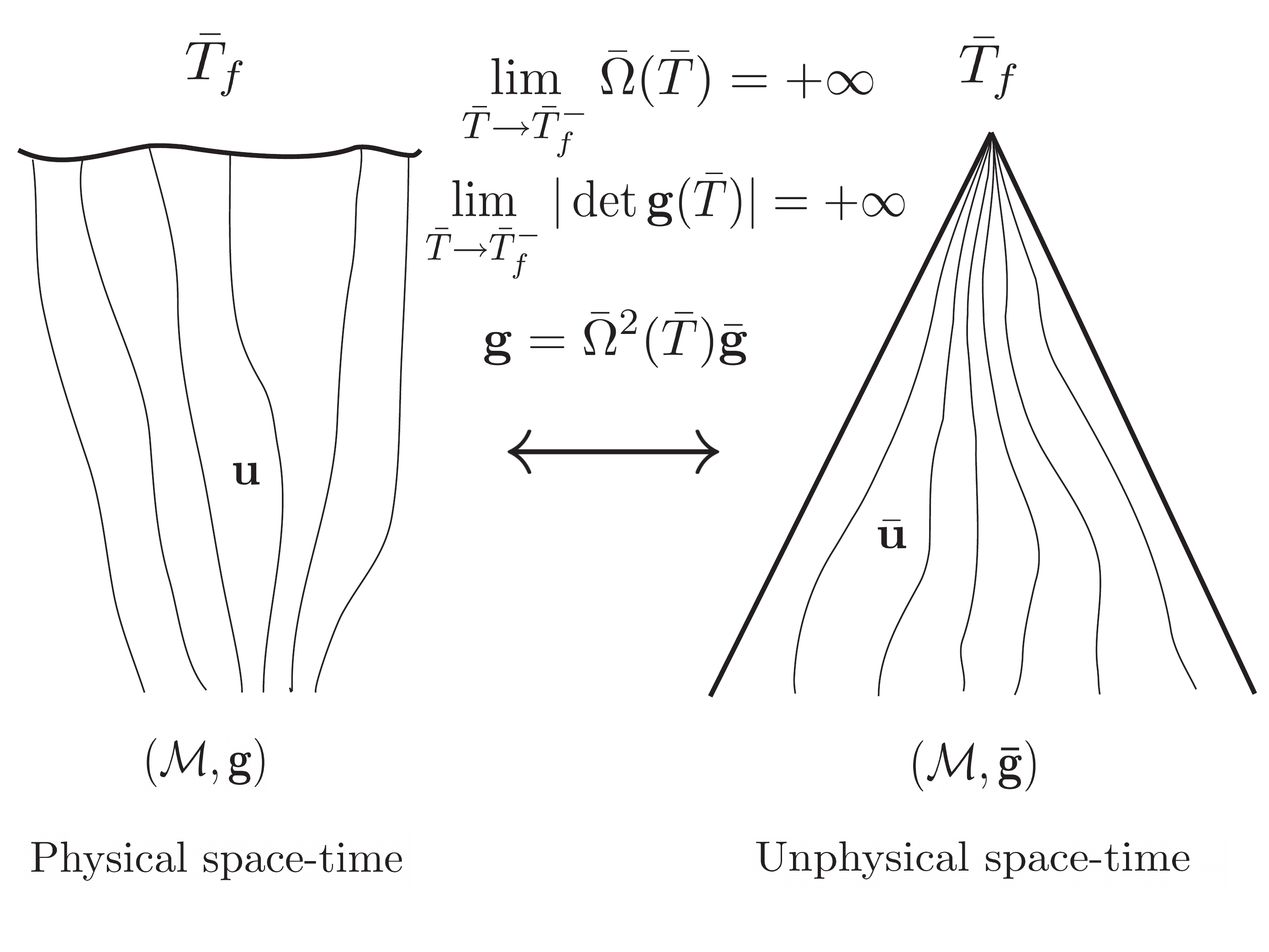}

%\end{figure}

%  	\inclimg{coolingfct}{}

%\begin{center}

%\vspace*{5cm}

\caption{{\footnotesize The schematic interpretation of the definition of an AFEU }}

%\end{center}

\end{figure}

The degeneracy of the unphysical metric $\bar{\mathbf{g}}$ produces a singularity in the unphysical space-time and, in fact,  forces that singularity as well as the physical AFS to be a \emph{deformationally strong singularity}, which highlights the final cosmological state. The physical expansion, $\theta$, of the cosmological fluid, furthermore, shows the desired behaviour at the cosmological futures, namely it contracts to caustics at the AFS and remains non-negative in the case of the AFEU. Finally, example cosmologies clearly show that the new definitions allow for the desired anisotropies to occur; the kinematics are no longer expansion dominated and the scalar $K$ takes values different from zero, which is essential for compatibility with \emph{quiescent cosmology} and the WCH \cite{ScottHoehn2}.

These results lead us to conjecture that the conjunction of the definition of an IPS with that of an AFS or an AFEU may provide the first complete mathematical formalisation of \emph{quiescent cosmology} in the sense that a cosmology satisfying these conditions possesses an evolution in accordance with this school of thought. 

%\begin{singlespace}

%\end{singlespace}

\end{document}